\documentclass[10pt,twocolumn,letterpaper]{article}

\usepackage[pagenumbers]{cvpr} 

\usepackage{graphicx}
\usepackage{amsmath}
\usepackage{amssymb}
\usepackage{booktabs}

\usepackage{algorithm}
\usepackage{algorithmic}
\usepackage{multirow}
\usepackage{makecell}
\usepackage{enumerate}
\usepackage{bm}

\usepackage[accsupp]{axessibility}  

\usepackage[pagebackref,breaklinks,colorlinks]{hyperref}

\usepackage[capitalize]{cleveref}
\crefname{section}{Sec.}{Secs.}
\Crefname{section}{Section}{Sections}
\Crefname{table}{Table}{Tables}
\crefname{table}{Tab.}{Tabs.}

\def\cvprPaperID{85} 
\def\confName{CVPR}
\def\confYear{2023}

\begin{document}

\title{Effective Ambiguity Attack Against Passport-based DNN Intellectual Property Protection Schemes through Fully Connected Layer Substitution}

\author{
 Yiming Chen\textsuperscript{\rm 1}, 
 Jinyu Tian\textsuperscript{\rm 2}, 
 Xiangyu Chen\textsuperscript{\rm 1,3}, and 
 Jiantao Zhou\textsuperscript{\rm 1,\footnotemark[2]} \\
 \textsuperscript{\rm 1}State Key Laboratory of Internet of Things for Smart City\\
 Department of Computer and Information Science, University of Macau\\
 \textsuperscript{\rm 2}Faculty of Innovation Engineering, Macau University of Science and Technology\\
 \textsuperscript{\rm 3}Shenzhen Institutes of Advanced Technology, Chinese Academy of Sciences\\
 {\tt\small \{yc17486, jtzhou\}@umac.mo}, {\tt\small jytian@must.edu.mo}, {\tt\small chxy95@gmail.com}
}

\maketitle
\renewcommand{\thefootnote}{\fnsymbol{footnote}} 
\footnotetext[2]{Corresponding author.} 

\begin{abstract}
   Since training a deep neural network (DNN) is costly, the well-trained deep models can be regarded as valuable intellectual property (IP)  assets. The IP protection associated with deep models has been receiving increasing attentions in recent years. Passport-based method, which replaces normalization layers with passport layers, has been one of the few protection solutions that are claimed to be secure against advanced attacks. In this work, we tackle the issue of evaluating the security of passport-based IP protection methods. We propose a novel and effective ambiguity attack against passport-based method, capable of successfully forging multiple valid passports with a small training dataset. This is accomplished by inserting a specially designed accessory block ahead of the passport parameters. Using less than 10\% of training data, with the forged passport, the model exhibits almost indistinguishable performance difference (less than 2\%) compared with that of the authorized passport. In addition, it is shown that our attack strategy can be readily generalized to attack other IP protection methods based on watermark embedding. Directions for potential remedy solutions are also given. 
\end{abstract}

\section{Introduction}
\label{sec:intro}

With the geometric growth of computing power of computational devices in recent decades, there have emerged many deep learning applications that have contributed to the human world such as super-resolution reconstruction \cite{chen2022activating, dong2015image, wang2018esrgan}, image inpainting \cite{9410590, yu2018generative, yan2018shift} and forgery detection \cite{9686650}. It usually costs many resources to develop new DNN models and developers will not tolerate the act of theft of their IP. The IP protection problem of deep models becomes more severe with the birth of Machine Learning as a Service (MLaaS) \cite{7424435}. Preventing the infringement behavior of deep models now emerges as a necessary concern when developing new algorithms and systems.     

Model watermark \cite{zhang2020model, li2021spread, 9413062, szyller2021dawn, namba2019robust} has been a popular method to protect the IP of DNN models. In the embedding process, the owners embed the secret signatures (watermarks), and then in the verification process, they can claim their ownership to the model by matching the extracted signatures with the original versions. The existing model watermark methods can be roughly divided into two categories\cite{9454280, NEURIPS2019_75455e06}: feature-based and trigger-based methods. Specifically, feature-based methods \cite{uchida2017embedding, darvish2019deepsigns, chen2019deepmarks, nagai2018digital} applied a regularizer to embed the secret watermark into the activation functions or model weights. Uchida \etal \cite{uchida2017embedding} proposed to use a regularizer to embed a watermark into the model weights.  Darvish \etal \cite{darvish2019deepsigns} embedded the fingerprints in the Probability Density Function of trainable weights instead. Aramoon \etal \cite{aramoon2021don} inserted the signature into the gradient of the cross-entropy loss function with respect to the inputs. In contrast, trigger-based methods make the output target respond to specific inputs. Along this line, Adi \etal \cite{adi2018turning} used the backdoor attack as a means to watermark the model. Merrer \etal \cite{le2020adversarial} designed a zero-bit watermarking algorithm that uses adversarial samples as watermarks to claim the ownership.  Zhang \etal \cite{zhang2018protecting} applied watermarks to images and then trained the network to output target labels when input images carry these watermarks.      
    
Despite the strength in retaining ownership of DNN models, most existing model watermark methods are shown to be vulnerable to the so-called ambiguity attack, in which the attacker manages to cast doubts on the ownership verification by crafting counterfeit (forged) watermarks\cite{9454280}. Recently, Fan \etal \cite{NEURIPS2019_75455e06} first designed a series of ambiguity attacks, which are effective in attacking DNN watermark methods. It was stated that for conventional watermark methods, a counterfeit watermark can be forged as along as the model performance is independent of the signature \cite{9454280}. Following this proposition, Fan \etal designed a passport layer through which the functionality of the model is controlled by the signature called \emph{passport}. However, Fan \etal encountered a heavy performance drop when batch normalization layers exist. To solve this problem, Zhang \etal \cite{zhang2020passport} added learnable affine transformations to the scale and bias factors. It was claimed that an attacker cannot find a substitute passport that maintains the model performance, which ensures the security of these passport-based methods against \textit{existing} ambiguity attacks.       
   
In this work, we aim to design an advanced ambiguity attack to the passport-based method, capable of generating valid substitute passports with only a small number of data. Here, valid substitute passports are defined as those leading to an indistinguishable model performance, but sufficiently different from the original authorized passports. Clearly, with such valid substitute passports, an attacker can claim the ownership of the model. To this end, we first experimentally justify the existence of multiple valid substitute passports. Noticing the fact that it is easy to localize the passport layers, we then propose our ambiguity attack by replacing passport layers with our designed two types of structures, namely \textbf{Individual Expanded Residual Block (IERB)} and \textbf{Collective Expanded Residual Block (CERB)}. Both structures are built in a way to encourage the significant changes of the parameters in the passport layers during the training, which could help us search for valid substitute passports. Benefiting from these two structures and assisting with a small amount training data, we can obtain valid substitute passports, and hence, defeat the passport-based methods which are the only type of method claimed to be immune to existing ambiguity attacks.            

Our major contributions can be summarized as follows:
\begin{itemize}
    \vspace{-3mm}
    \item We propose a novel and effective ambiguity attack against the passport-based IP protection schemes. With less than 10\% of training data, our ambiguity attack on passport-layer protected model can restore the functionality of the model with a less than 2\% performance gap from the original accuracy. 
    \vspace{-3mm}
    \item  We design two novel structures for replacing the passport layers, based on the multi-layer perceptron (MLP) and skip connection to assist with our ambiguity attack for searching valid substitute passports with a small amount of training data. 
    \vspace{-3mm}
    \item Experiments on both overlapping (attacker's training dataset is part of the original training dataset) and non-overlapping datasets (attacker's dataset and the original one come from the same source but no overlap exists), and on different network structures have proved the effectiveness of our ambiguity attack.
    \vspace{-3mm}
    \item Our attack method can be readily generalized to attack other DNN watermark methods \cite{darvish2019deepsigns, liu2021watermarking, uchida2017embedding}.
\end{itemize}

\section{Related Works}

DNN watermark methods have been popular solutions for DNN model IP protection. However, these techniques might still be vulnerable to flagrant infringement from notorious adversaries. In this section, we review the two types of representative attack methods, namely, removal attack \cite{chen2021refit, guo2020fine, chen2019leveraging, aiken2021neural, liu2021removing, 2205.00199} and ambiguity attack \cite{NEURIPS2019_75455e06, zhang2020passport, 9454280}, along with the passport-based method attempting to defend against ambiguity attacks \cite{9454280}.


\textbf{Removal Attack}: This type of attack tries to remove the watermark from the protected model, malfunctioning the ownership verification mechanism. Along this line, many fine-tuning based methods have been proposed. Chen \etal \cite{chen2021refit} combined a redesigned elastic weight consolidation algorithm and unlabeled data augmentation to achieve unified model watermark removal with limited data. Guo \etal \cite{guo2020fine} used a dataset transformation method called PST (Pattern embedding and Spatial-level Transformation) to preprocess the data before fine-tuning. Chen \etal \cite{chen2019leveraging} utilized auxiliary unlabeled data to decrease the amount of labeled training data required for effective watermark removal. Aiken \etal \cite{aiken2021neural} 
provided a three-stage scheme to remove backdoor-based watermarks by exploiting another trigger-free dataset from the same domain. Liu \etal \cite{liu2021removing} designed a framework to remove backdoor-based watermarks, in which a data augmentation was proposed to imitate the behavior of the backdoor triggers. Yan \etal \cite{2205.00199} attempted to break the passport-based method by scaling the neurons and flipping the signs of parameters. However, this method assumed that the authorized passports are available to the attacker, which is not realistic in practice. Also, these aforementioned attack methods only enable the attackers to remove the watermarks, while unable to claim the ownership.   

\textbf{Ambiguity Attack}: Another more threatening attack is the ambiguity attack, where the attacker can forge another substitute watermark to claim the model ownership. The concept of ambiguity attack originally appeared in image watermark community \cite{LOUKHAOUKHA2017359, li2006zero}, and recently has been extended to the DNN watermark methods. The pioneering work was conducted by Fan \etal in \cite{NEURIPS2019_75455e06}, which pointed out the vulnerability of Uchida's watermark method \cite{uchida2017embedding} under the ambiguity attack. They also showed that the same weakness of  Adi's DNN watermark method \cite{adi2018turning} exists, by proving that another trigger can be optimized exclusively to cause the same model response as the original one.               

\textbf{Passport-based method}: Passport-based method was originally proposed by Fan \etal \cite{9454280} as a remedy enabling DNN watermark methods to defeat the ambiguity attack. This is achieved by replacing the traditional normalization layer with the so-called passport layer, whose difference mainly lies in  how the affine factors are obtained. In passport layer, the scale factor $\gamma$ and bias factor $\beta$ are computed with the passport as follows:
\begin{equation}
    \gamma = Avg(\mathbf{W}_{conv} * \mathbf{s}_{\gamma}), \beta = Avg(\mathbf{W}_{conv} * \mathbf{s}_{\beta}),
\end{equation} where $\mathbf{s} = \left\{\mathbf{s}_{\gamma}, \mathbf{s}_{\beta}\right\}$ is called the passport, $\mathbf{W}_{conv}$ is the convolutional layer weight before this layer, and $Avg(\cdot)$ represents the average pooling function.          

To embed the passport $\mathbf{s}$ into the model, the network $\mathbb{N}_p$ is optimized on the training set $\mathcal{D}=\left\{(x_i, y_i)\right\}_{i=1}^N$, where $x_i$ is the input and $y_i$ is the corresponding label, using the following loss:

\vspace{-4mm}

\begin{equation}
    \mathcal{L} = \mathcal{L}_r(\mathbb{N}_p[\mathbf{W}, \mathbf{s}](x_i, y_i)) + \alpha \mathcal{L}_{sign}(\mathrm{sgn} (\gamma), \mathbf{b}),
\end{equation}
\noindent
where $\mathbb{N}_p[\mathbf{W}, \mathbf{s}]$ denotes the model $\mathbb{N}$ with the weight $\mathbf{W}$ and the passport $\mathbf{s}$. The first term $\mathcal{L}_r(\cdot)$ is the classification loss, and the second term $\mathcal{L}_{sign}(\cdot)$ is a sign loss regularization constraining the sign sequence of scale factors to be a predefined \emph{C}-bit signature $\mathbf{b}=\left\{b_1,...,b_C\right\} \in \left\{-1, 1\right\}^C$. Here, $\mathrm{sgn}(\cdot)$ is the sign function. 

Denote $\mathcal{Q}(\mathbb{N}_p[\mathbf{W}, \mathbf{s}])$ as the inference performance of model $\mathbb{N}_p$ with the passport $\mathbf{s}$. The verification process mainly relies on the assumption that the performance $\mathcal{Q}(\mathbb{N}_p[\mathbf{W}, \mathbf{s}])$ seriously deteriorates when an unauthorized passport is presented. Hence, the model ownership using passport-based method is conducted by matching the signs of scale factors $\mathrm{sgn} (\gamma)$ with the predefined signature $\mathbf{b}$ and checking the DNN model inference performance.

\section{Proposed Ambiguity Attack}

Though the passport-based methods \cite{zhang2020passport, 9454280} have been claimed to be immune to the existing ambiguity attacks \cite{NEURIPS2019_75455e06}, they do not preclude the existence of more advanced ambiguity attacks. In this work, we cast doubt on the security of the passport-based methods, and propose a simple yet effective ambiguity attack, capable of crafting valid substitute passports. Before diving into the detailed design of our ambiguity attack, let us clarify our attack goal and the information assumed to be available to the attacker.


\textbf{Attack Goal}: Essentially, the attack goal is to generate substitute passport  $\mathbf{s}_t$, with which the model $\mathbb{N}_p[\mathbf{W}, \mathbf{s}_t]$ has indistinguishable performance with that of applying the authorized passport $\mathbf{s}_o$, i.e., $\mathcal{Q}(\mathbb{N}_p[\mathbf{W}, \mathbf{s}_o])$. Specifically, following the ambiguity attack setting in \cite{9454280}, we assume that the attacker has access to the model weights $\mathbf{W}$ except from the passport. Note that, without the passport, the affine factors are missing in the normalization layer, through which the attacker can easily localize the passport layers. Also, the attacker is allowed to have a small number of training data, namely, $\mathcal{D}_s=\left\{(\hat{x}_i, \hat{y}_i)\right\}_{i=1}^n$, where $n << N$. Formally, we have the following \textbf{Definition 1}, explicitly explaining the successful ambiguity attack on passport-based method.                      
	
\begin{figure}[t] \centering
	\includegraphics[width=0.47\textwidth]{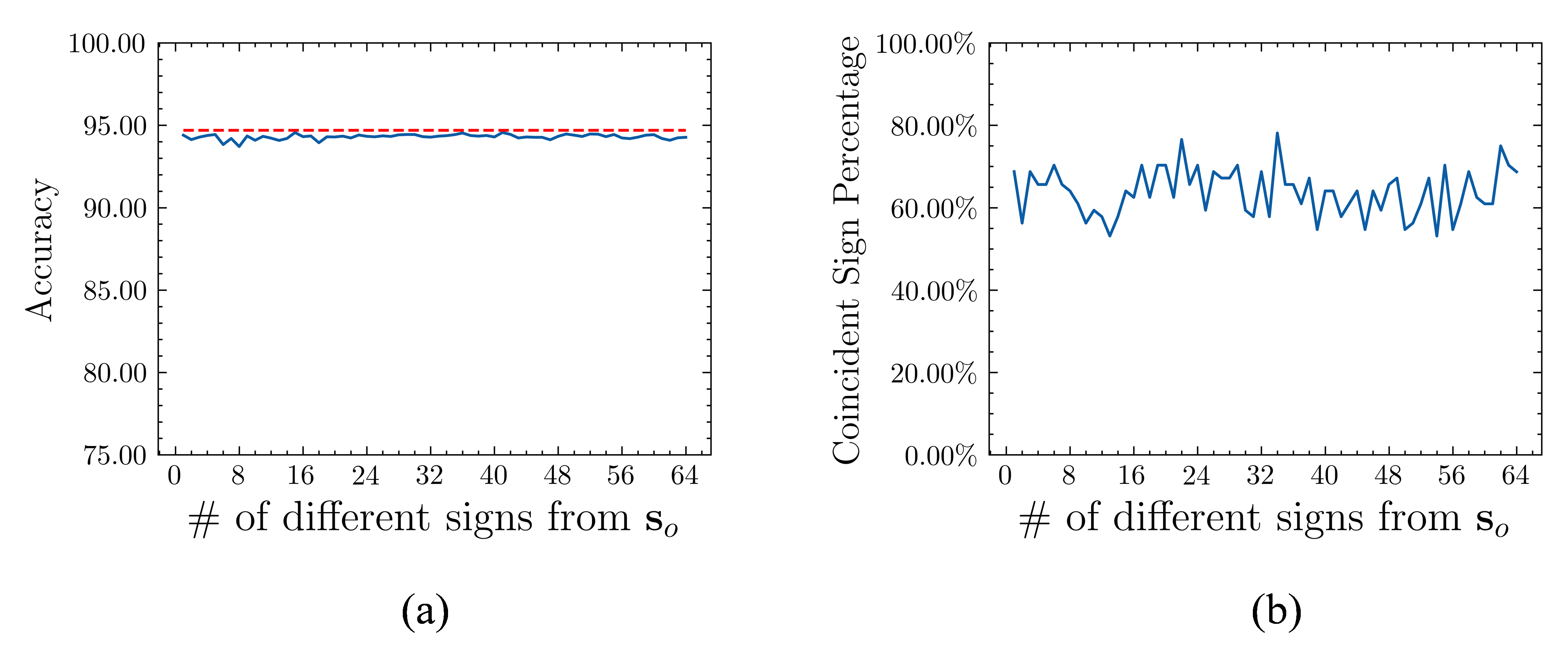}
	\vspace{-2mm}
	\caption{Valid substitute passports obtained by initializing scale factors with different signs from the authorized $\mathbf{s}_o$ and retraining. Horizontal axis denotes the number of different signs from $\mathbf{s}_o$.}
	\vspace{-3mm}
	\label{flip}    
\end{figure}

\begin{figure*}[t]
\centering
\includegraphics[width=0.9\textwidth]{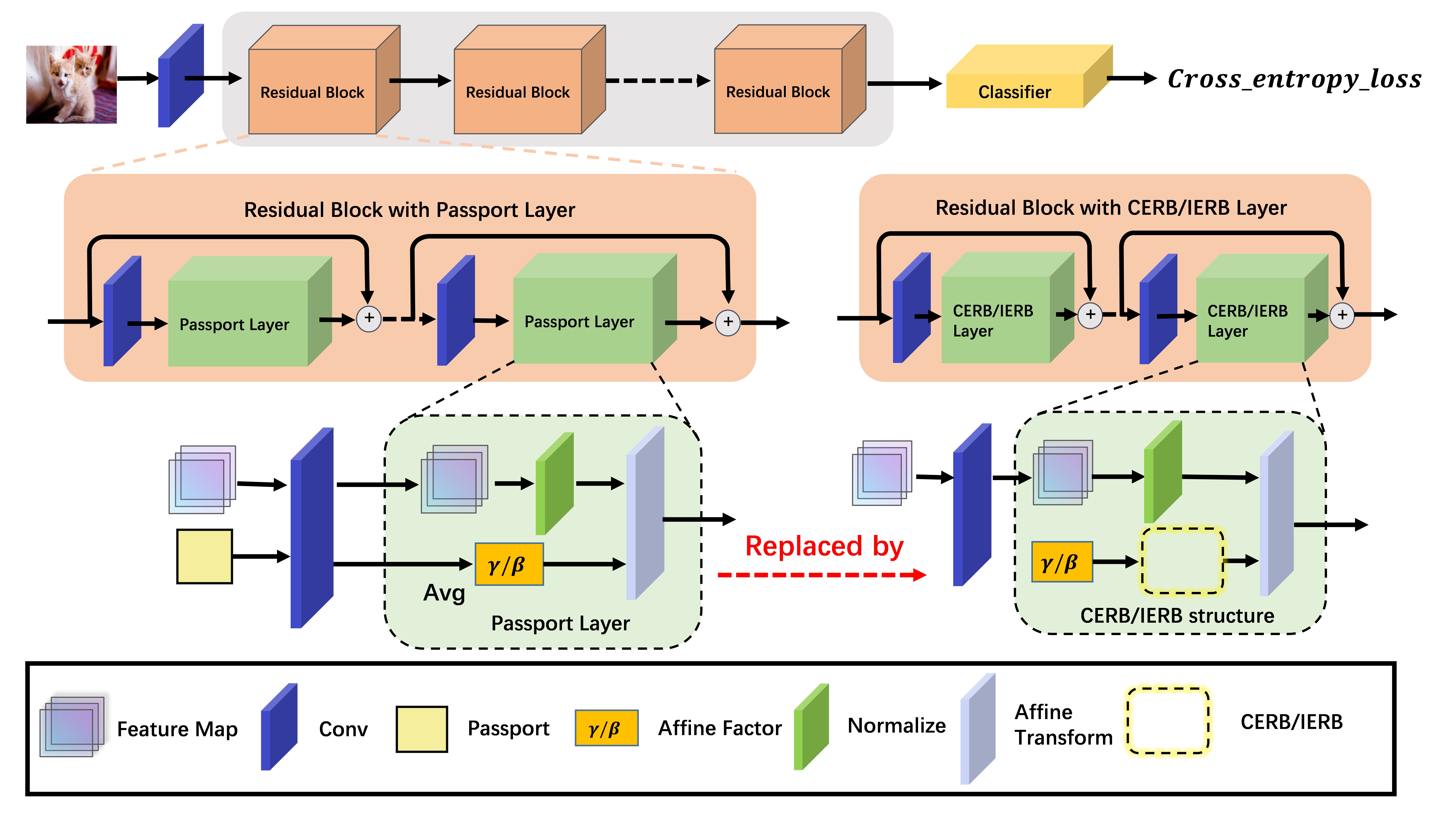}
\vspace{-2mm}
\caption{The structures of the ResNet18 with passport layer and our designed CERB/IERB structures.}
\vspace{-3mm}
\label{overview}
\end{figure*}

\textbf{Definition 1:} An ambiguity attack on passport layer protected model is successful, if
\begin{enumerate}[I)]
	\item DNN inference accuracy with the substitute passport $\mathbf{s}_t$ is close to that with the authorized passport $\mathbf{s}_o$, i.e.,   

        \vspace{-2mm}
 
	\begin{equation}
	\label{success-acc-indistinguish}
	    \Big | \mathcal{Q}(\mathbb{N}_p[\mathbf{W}, \mathbf{s}_t]) - \mathcal{Q}(\mathbb{N}_p[\mathbf{W}, \mathbf{s}_o]) \Big | < \epsilon.
	\end{equation}
	
	\item Dissimilarity between the substitute passport $\mathbf{s}_t$ and the authorized $\mathbf{s}_o$ should be large enough, i.e.: 

    \vspace{-2mm}
 
    \begin{equation}
    \label{success-passport-difference}
    \Big | \mathbf{s}_t - \mathbf{s}_o \Big | > \delta.
    \end{equation}
\end{enumerate}

Before presenting the details of our ambiguity attack, let us first justify the existence of multiple valid substitute passports satisfying the conditions given in (\ref{success-acc-indistinguish}) and (\ref{success-passport-difference}). To this end, we here adopt an experimental approach and leave the theoretical justification in our future work. Assume now that we are given the complete training data $\mathcal{D}$, though this large amount of data are not required when launching the attack. We initialize scale factors with different combinations of $\{+1,-1\}$ and fine-tune the model based on $\mathcal{D}$. In Fig. \ref{flip}, we give the experimental results on ResNet18 trained on the CIFAR10 dataset, where the passport layer is placed after the first convolutional layer and the length of the scale factor is 64. The model with the authorized passport leads to an inference performance of 94.70\%. As can be seen, the accuracy of the models after retraining is still close to 94.70\% (red line, see Fig. \ref{flip}(a)). More importantly, the signs of retained affine factors only have low coincidence rate (around 60\%, see Fig. \ref{flip}(b)) with the original ones, implying that retained affine factors differ significantly from the authorized affine factors. Therefore, these retrained affine factors could simultaneously satisfy the conditions  (\ref{success-acc-indistinguish}) and (\ref{success-passport-difference}), and hence are valid substitute passports. 

Though the existence of substitute passports has been justified, the difficulty of directly optimizing a passport remains unsolved for very limited number of training data. Clearly, in practical attacks, the attacker is only allowed to have access to very limited data; otherwise, he can retrain the entire model. Fan \etal \cite{9454280} ascribed the robustness of passport-based method against fine-tuning the scale factors to the lazy-to-flip property, with which the scale factors are rarely updated to cross the zero during the training. 

To overcome the lazy-to-flip property for the scale factors, we attempt to add a trainable block after it, encouraging scale factors to go across the zeros by non-linear transformations. For efficiency, we adopt MLP for designing the trainable blocks. Following this line, we design two structures namely IERB and CERB to replace the passport layer.           

\vspace{-4mm}

\paragraph{The overall structure:} Motivated by the above observations, we now design a novel ambiguity attack on passport-based method. The overall structure is given in Fig. \ref{overview}, where we use the ResNet18 embedded with passport layers for the illustration. Specifically,  the protected ResNet18 is comprised of multiple residual blocks, each containing several convolutional layers followed by the passport layer. As aforementioned, the locations of the passport layers can be easily determined. We can then insert either IERB or CERB structure into these locations. In our structure, the scale factor $\gamma$ is the output of the IERB/CERB. For the bias factor $\beta$, we do not add our new structure; but instead we optimize it directly. We are now ready to introduce the details of CERB and IERB structures. After that, we will give the complete algorithm of our ambiguity attack.             
	
\subsection{Individually Expanded Residual Block (IERB)}
\label{IERB-intro}

\begin{figure}[!t]
        \centering 
        \includegraphics[width=0.45\textwidth]{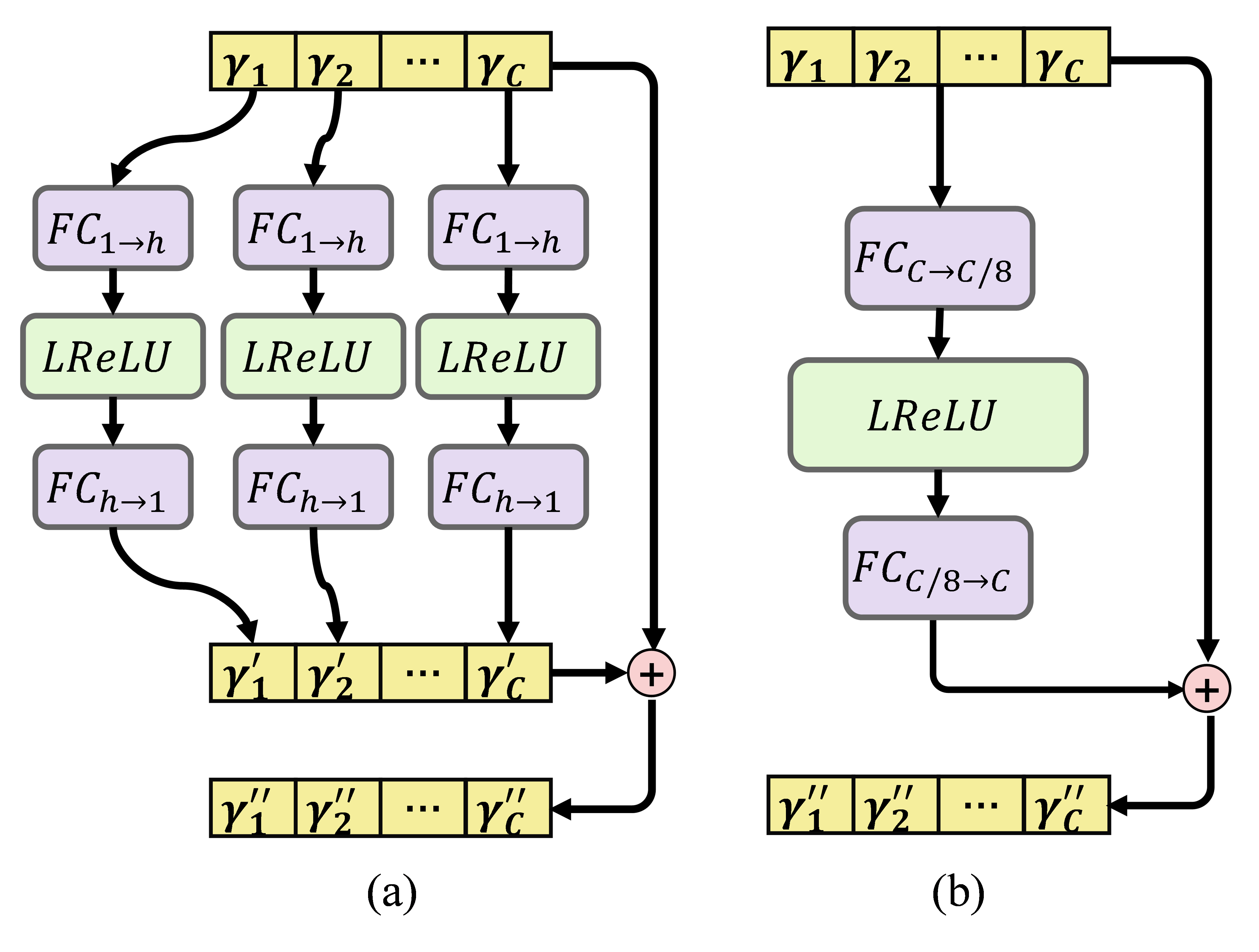}
        \vspace{-2mm}
        \caption{Details of (a) IERB and (b) CERB.}
        \vspace{-3mm}
        \label{Block Structure}
    \end{figure}

In this subsection, we present the details of the IERB block. As showed in Fig. \ref{Block Structure} (a), the $i$-th scale factor $\gamma^{l}_{i}$ in the $l$-th passport layer is transformed by a Two-Layer Perceptron (TLP) with FC-LeakyReLU-FC structure where the FC refers to fully connected layer. For the simplicity of notations, we omit the superscript in $\gamma^{l}_{i}$ in the sequel. The output of this TLP is then used for the affine transformation later. The first linear layer $FC_{1\rightarrow h}$ with learnable parameters maps the scale factor into a $h$-dimensional vector, while the second linear layer $FC_{h\rightarrow 1}$ maps the dimension back to 1, where $h$ is the hidden layer size. A larger $h$ brings a larger optimization space in the hidden layer; but could add burden to the training process. We empirically set $h=10$, striking a good balance between these two factors. In addition, a skip connection is added aside with the TLP structure to facilitate the back-propagation. Benefiting from the expanded parameter space brought by the TLP structure, the output scale factor would more likely go across the zero during the updating. This helps the attacker search for an appropriate $\mathbf{s}_t$ satisfying (\ref{success-passport-difference}).           

Let $\{\mathbf{W}_{i, j}, \mathbf{b}_{i, j}\}$ be the weights of the $j$-th linear layer connected after the $i$-th scale factor, where $i \in \{1,2,\cdots, C\}$ and $j\in \{1,2\}$. In IERB, the output of the first linear layer taking the $i$-th scale factor as input is: 

\vspace{-3mm}

\begin{equation}
	\label{IERB-eq}
    \gamma '_i= \mathbf{W}_{i, 1}^T * \gamma_i+ \mathbf{b}_{i, 1}^T,
\end{equation} where $*$ denotes the matrix multiplication operator. After the activation function LeakyReLU, the output of the second linear layer with the shortcut can be expressed as: 

\vspace{-3mm}

\begin{equation}
    \gamma''_i = \mathbf{W}_{i, 2}^{T} * ( LReLU(\gamma'_i)) + \mathbf{b}_{i, 2}^{T} + \gamma_{i}.
\end{equation} 


\subsection{Collective Expanded Residual Block (CERB)}
\label{CERB-intro}
	
The IERB discussed above handles each scale factor individually, but may ignore the inter-channel relationship among scale factors. Instead of using isolated propagation paths for each scale factor, we now attempt to enhance the inter-channel relationship and optimize the scale factors in a collective manner. As can be observed from Fig. \ref{Block Structure} (b), we choose to use a unified TLP structure to deal with all the $C$ scale factors as a whole.  The output of the TLP structure is then summed with the input scale factors to get the updated ones.  Different from the IERB, each activation in the hidden layer is now related to all the input scale factors, and the fully connected structure ensures the exploitation of the inter-channel information. Mathematically, the output scale factors can be computed by:       

\vspace{-5mm}

\begin{equation}
    \label{CERB-eq}
	\mathbf{\gamma}'' = ( \mathbf{W}_{2}^{T} * ( LReLU( \mathbf{W}_{1}^{T} * \gamma + \mathbf{b}_{1}^{T} ) ) + \mathbf{b}_{2}^{T} ) + \mathbf{\gamma},
\end{equation}
where $\mathbf{W}_{1}$ and $\mathbf{W}_{2}$ represent the parameters in the first and second linear layers. Here we set the hidden layer size to be $C/8$, where $C$ is the number of channels in this layer.

\subsection{Algorithm of Our Ambiguity Attack}

\newlength{\textfloatsepsave} 
\setlength{\textfloatsepsave}{\textfloatsep}

\setlength{\textfloatsep}{2pt}
\begin{algorithm}[t]
\caption{Proposed Ambiguity Attack}
\label{alg:algorithm}
\textbf{Input}: Protected network $\mathbb{N}_p[\mathbf{W}]$ with passport layers excluded; training dataset $\mathcal{D}_s=\left\{(\hat{x}_i, \hat{y}_i)\right\}_{i=1}^n$;  checkpoint \emph{state\_dict};  training epoch $M$. 

\textbf{Output}: substitute scale and bias factors $\gamma''$, $\beta$.

\begin{algorithmic}[1] 
\STATE Use normalization layers in the locations of the passport layers.   
\STATE Insert CERB/IERB structures after the scale factors of these normalization layers.
\STATE Load weight $\mathbf{W}$ of $\mathbb{N}_p[\mathbf{W}]$ from \emph{state\_dict}. 
\STATE Initialize normalization layer weights $\gamma$ and $\beta$ with $\mathbf{1}$ and $\mathbf{0}$, respectively.  
\FOR{epoch = 1 to $M$}
\FOR{\emph{minibatch} $(\hat{x}_i,\hat{y}_i) \subset \mathcal{D}_s$}
\FOR{each normalization layer with IERB/CERB}
\STATE $\gamma ''$ = CERB/IERB($\gamma, \mathbf{W}_{\gamma}$). \hspace*{\fill} \mbox{$\triangleright$ Eq. (\ref{IERB-eq}/\ref{CERB-eq})}. 
\STATE Use $\gamma ''$ and $\beta$ for affine transformation in normalization layer.
\ENDFOR
\STATE \emph{loss} = \emph{cross\_entropy}($\mathbb{N}_p[\mathbf{W}](\hat{x}_i), \hat{y}_i$).
\STATE Update $\gamma$, $\beta$ and $\mathbf{W}_{\gamma}$. \mbox{$\triangleright$ Eq. (\ref{update-params})}.
\ENDFOR
\ENDFOR
\end{algorithmic}
\end{algorithm}

With these newly proposed structures, we can summarize our ambiguity attack strategy on the passport-based method in Algorithm \ref{alg:algorithm}. For simplicity, we use $\mathbf{W}_\gamma$ to represent all the parameters in IERB or CERB. Let us briefly explain our workflow for better understanding. In Algorithm \ref{alg:algorithm}, lines 1$\sim$4 are devoted to the model loading, normalization layer substitution, CERB/IERB insertion, and parameter initialization. The $\gamma$, $\beta$ and $\mathbf{W}_{\gamma}$ are then updated in lines 5$\sim$14 using the gradients with respect to each parameter by back-propagating the cross entropy \emph{loss}:

\vspace{-2mm}

\begin{equation}
\label{update-params}
\begin{aligned}
     \gamma &= \gamma - \nabla_{\gamma}\emph{loss}, \\
    \beta &= \beta - \nabla_{\beta}\emph{loss}, \\
    \mathbf{W}_{\gamma} &= \mathbf{W}_{\gamma} - \nabla_{\mathbf{W}_{\gamma}} \emph{loss}.
\end{aligned}
\end{equation}

Eventually, the algorithm outputs the substitute scale and bias factors $\gamma''$ and $\beta$. 

\section{Experimental Results}
In this section, we evaluate the effectiveness of our ambiguity attack from different perspectives. Before presenting the detailed results, let us clarify the experimental settings including the datasets, target models, and the evaluation metrics.


\textbf{Dataset}: Four image classification datasets: CIFAR10 \cite{krizhevsky2009learning}, CIFAR100 \cite{krizhevsky2009learning}, Caltech-101 \cite{fei2004learning}, and Caltech-256 \cite{griffin2007caltech}, are selected. Unless stated, the dataset used in the attack process only accounts for 10\% \textit{at maximum} of the full training set and does not overlap with the test set. 

\setlength{\textfloatsep}{\textfloatsepsave}

\textbf{DNN architectures}: Three DNN architectures, AlexNet \cite{krizhevsky2012imagenet}, ResNet-18 \cite{he2016deep} and Wide-Residual Network \cite{zagoruyko2016wide} are used in our experiments, following the tradition of passport-based works \cite{NEURIPS2019_75455e06, zhang2020passport, 9454280}. To demonstrate that our attack strategy remains effective for different number of passport layers, we perform the experiments on AlexNet and ResNet18, with at most 5 and 18 passport layers, respectively. For notation convenience, we use a suffix to determine the indices of the passport layers. For instance, AlexNet-4 denotes the AlexNet with the first 4 normalization layers replaced by the passport layers, and AlexNet-last3 represents the version in which the last 3 normalization layers are replaced by the passport layers.

\textbf{Evaluation metrics}: Prediction accuracy (ACC) of the model is a natural choice for evaluating the attack effectiveness, since a successfully forged passport is expected to achieve similar performance as an authorized passport (see Definition 1). Another metric considered is the bit dissimilarity rate (BDR) of the signature derived from the forged passport, with respect to the authorized one. Specifically, let $\gamma'$ and $\gamma^{o}$ be the forged and authorized scale factors, respectively. Note that the sign of $\gamma^{o}$ is used as the signature. The BDR is then defined as:                 

\vspace{-2mm}

\begin{equation}
    BDR = \dfrac{1}{C}\sum\limits_{i=1}\limits^{C}\mathcal{I}\Big[ \mathrm{sgn} (\gamma'_i) \neq \mathrm{sgn} (\gamma^o_i) \Big],
\end{equation}

\noindent
where the indicator function $\mathcal{I}[\cdot]$ returns 1 when the condition is true and 0 otherwise. A high BDR also implies the large dissimilarity with the authorized passport, which in turn indicates a better performance of the ambiguity attack.      

\subsection{Attack performance using IERB/CERB}
\label{direct-attack-result-section}
We now give the details on the effectiveness of our ambiguity attack by replacing the passport layers of the protected model with our proposed IERB or CERB blocks. All protected models with passport layers are trained over the training dataset of CIFAR10. We randomly select 10\% of examples from this dataset for launching the ambiguity attack, i.e., training the IERB or CERB blocks.

The attack results on AlexNet and ResNet18 are reported in Table \ref{direct-attack-result}. Since our ambiguity attack is the \textit{first one} attacking passport-based methods, there are no comparative algorithms available so far. It can be seen that, by using the proposed CERB, our ambiguity attack is capable of achieving high ACC and BDR values for all the settings. Specifically, for AlexNet, the ACC gap is less than 1\%, compared with the case of the authorized passport. Also, the BDR can be as large as 80.30\%. For the more complex ResNet18, similar observations can be obtained, where the ACC gap is still less than 3\%, and the BDR could approach 50\%. We even encounter several cases (AlexNet-1 and ResNet18-1), in which the attacked models perform even better than the original ones. These results imply that our proposed ambiguity attack is successful in finding the valid substitute passports with a small number of training data.  

As a comparison, we also present the results of the ambiguity attack with the IERB structure. It can be observed that this variant attack is still somewhat effective with reasonably high ACC and BDR values; but the attack performance is much inferior to the attack with CERB, especially when there are many passport layers. We attribute this phenomenon to the capability of CERB in exploiting the inter-channel relationship. For IERB, it is likely to optimize towards the original sign of the scale factor, which naturally results in similar signatures (small BDR) to the original passports. In CERB, however, the collective optimization mechanism explores more possibly optimal pairs of the sign of scale factors, potentially leading to larger BDR of signature and higher ACC.

\begin{table}[t]
\centering
\scalebox{0.85}{
\begin{tabular}{l|c|cc|cc}
\toprule
\cmidrule{1-6}
\multirow{2}{*}{Name} & Original & \multicolumn{2}{l|}{IERB}      & \multicolumn{2}{l}{CERB}      \\ \cmidrule{3-6} 
                      &  ACC                       & \multicolumn{1}{l|}{ACC} & BDR & \multicolumn{1}{l|}{ACC} & BDR \\ \midrule
AlexNet-1            &  90.20                  & \multicolumn{1}{l|}{89.54}    &    16.00  & \multicolumn{1}{l|}{90.26}    &   24.67  \\ 
AlexNet-2            &  88.73                  & \multicolumn{1}{l|}{86.01}    &   9.76   & \multicolumn{1}{l|}{88.37}    &   46.10  \\ 
AlexNet-3            &  90.08                  & \multicolumn{1}{l|}{87.48}    &  12.96    & \multicolumn{1}{l|}{88.95}    &  41.41   \\ 
AlexNet-4            &  88.25                  & \multicolumn{1}{l|}{86.52}    &    16.96  & \multicolumn{1}{l|}{88.01}    &   79.80 \\ 
AlexNet-5            &  88.88                  & \multicolumn{1}{l|}{82.49}    &    27.95  & \multicolumn{1}{l|}{88.20}    & 80.30   \\

\hline
\noalign{\vskip\doublerulesep\vskip-\arrayrulewidth} 
\hline

ResNet18-1            &  94.89                  & \multicolumn{1}{l|}{93.85}    &   3.13  & \multicolumn{1}{l|}{94.94}    &   25.00  \\ 
ResNet18-3            &  94.80                  & \multicolumn{1}{l|}{93.46}    &   26.57  & \multicolumn{1}{l|}{94.22}    &  50.0   \\ 
ResNet18-4            &  94.67                  & \multicolumn{1}{l|}{92.92}    &   24.61  & \multicolumn{1}{l|}{94.09}    &  33.98   \\ 
ResNet18-5            &  94.39                  & \multicolumn{1}{l|}{90.04}    &   26.05  & \multicolumn{1}{l|}{93.75}    &    34.69 \\ 
ResNet18-6            &  94.45                  & \multicolumn{1}{l|}{93.48}    &   30.13  & \multicolumn{1}{l|}{92.44}    &    43.31 \\ 
ResNet18-8            &  94.41                  & \multicolumn{1}{l|}{91.22}    &   44.04  & \multicolumn{1}{l|}{92.66}    &  47.87   \\ 
ResNet18-10           &  94.30                  & \multicolumn{1}{l|}{88.96}    &   44.59  & \multicolumn{1}{l|}{92.87}    &  38.55   \\ 
ResNet18-12           &  94.96                  & \multicolumn{1}{l|}{87.36}    &   40.90  & \multicolumn{1}{l|}{92.34}    &  37.19   \\ 
ResNet18-14           &  94.68                  & \multicolumn{1}{l|}{86.63}    &   35.48  & \multicolumn{1}{l|}{92.18}    &  43.04   \\ 
ResNet18-18           &  94.81                  & \multicolumn{1}{l|}{83.92}    &   36.52  & \multicolumn{1}{l|}{92.11}    &  46.79   \\ 

\bottomrule
\end{tabular}}
\caption{ACC (\%) and BDR (\%) performance of ambiguity attacks using IERB/CERB structure for AlexNet and ResNet18 with different number of passport layers on CIFAR10.}
\vspace{-3mm}
\label{direct-attack-result}
\end{table}

\begin{table}[htbp]
\centering
\scalebox{0.8}{
\begin{tabular}{l|c|llll}
\toprule
\cmidrule{1-6}
Dataset-Net                        &Original & DataSize   & Plain & CERB  & IERB  \\
\midrule
\multirow{5}{*}{\makecell[c]{CIFAR10 \\ AlexNet}}   & \multirow{5}{*}{90.20}  & 5000 (10\%) & 83.62 & 88.16 & 87.30 \\
                                   & & 2500 (5\%)  & 73.82 & 86.70 & 78.91 \\
                                   & & 2000 (4\%)  & 73.48 & 86.30 & 76.73 \\
                                   & & 1500 (3\%)  & 72.09 & 86.78 & 71.38 \\
                                   & & 1000 (2\%)  & 64.82 & 84.22 & 63.71 \\
\hline
\noalign{\vskip\doublerulesep\vskip-\arrayrulewidth} 
\hline
\multirow{5}{*}{\makecell[c]{CIFAR100 \\ ResNet18}} & \multirow{5}{*}{75.05}  &5000 (10\%) & 70.73 & 73.26 & 70.51 \\
                                   & &3000 (6\%)  & 69.14 & 72.26 & 69.35 \\
                                   & &2000 (4\%)  & 62.81 & 71.42 & 68.89 \\
                                   & &1500 (3\%)  & 62.86 & 71.69 & 68.50 \\
                                   & &1000 (2\%)  & 53.91 & 70.12 & 68.44 \\ \bottomrule
\end{tabular}}
\caption{ACC (\%) comparison with Plain attack for CIFAR10-AlexNet and CIFAR100-ResNet18. }
\vspace{-1mm}
\label{result-vs-plain}
\end{table}

\begin{table}[htbp]
\scalebox{0.8}{
\begin{tabular}{c|l|c|c|c|c}
\toprule
\cmidrule{1-6}
Dataset                     & Passport Layer & Original & Plain & CERB & IERB \\ \hline
\multirow{4}{*}{Caltech101} & Last 3 layers  & 72.54 & 71.01 & 72.26 & 67.79\\ \cline{2-6} 
                            & Last 5 layers  & 68.30 & 59.89 & 66.78 & 60.73\\ \cline{2-6} 
                            & Last 8 layers  & 72.49 & 59.94 & 69.38 & 59.32\\ \cline{2-6} 
                            & Last 10 layers & 70.90 & 57.17 & 66.89 & 57.06\\ \hline \hline
\multirow{4}{*}{Caltech256} & Last 5 layers  & 54.34 & 38.64 & 52.23 & 51.02\\ \cline{2-6} 
                            & Last 7 layers  & 53.95 & 36.72 & 49.55 & 49.36\\ \cline{2-6} 
                            & Last 10 layers & 55.43 & 24.93 & 47.85 & 44.19\\ 
\bottomrule
\end{tabular}}
\caption{ACC (\%) comparison with Plain attack on Caltech-101 and Caltech-256, where 10\% training data are used. Here the network is fixed to be ResNet18.}
\vspace{-3mm}
\label{result-caltech}
\end{table}

\begin{table}[htp]
\centering
\scalebox{0.9}{
\begin{tabular}{l|cc|cc}
\toprule
 \cmidrule{1-5}
\multirow{2}{*}{DataSize} & \multicolumn{2}{c|}{IERB}      & \multicolumn{2}{c}{CERB}      \\ \cmidrule{2-5} 
                               & \multicolumn{1}{l|}{ACC} & BDR & \multicolumn{1}{l|}{ACC} & BDR \\ \hline
10000 (25\%)                    & \multicolumn{1}{c|}{90.47}    &  40.24   & \multicolumn{1}{c|}{92.17}    &  91.77   \\ \hline
5000 (12.5\%)                               & \multicolumn{1}{c|}{88.26}    &  42.82    & \multicolumn{1}{c|}{91.05}    &   92.82      \\ \hline
4000 (10\%)                               & \multicolumn{1}{c|}{87.43}    &  47.71   & \multicolumn{1}{c|}{90.07}    & 92.61 \\ \hline
3000 (7.5\%)                               & \multicolumn{1}{c|}{86.30}    &  48.34   & \multicolumn{1}{c|}{89.49}    &  93.65  \\ \hline
2500 (6.25\%)                               & \multicolumn{1}{c|}{86.23}    &  48.96   & \multicolumn{1}{c|}{87.00}    &  93.44   \\ \hline
2000 (5\%)                               & \multicolumn{1}{c|}{85.26}    &   49.20  & \multicolumn{1}{c|}{85.67}    &  93.44   \\ \hline
\end{tabular}}
\caption{ACC (\%) and BDR (\%) performance of ambiguity attacks with IERB and CERB using non-overlapping CIFAR10 dataset.}
\vspace{-4mm}
\label{ood}
\end{table}

To further show the effectiveness of our proposed attack strategy, we now compare the attack performance of the models with and without CERB/IERB structures. As will be clear soon, directly updating the affine factors cannot retrieve valid passports, especially when the available data is rather limited. To this end, we embed 5 passport layers in AlexNet and 10 passport layers in ResNet18, and train them on CIFAR10 and CIFAR100, respectively. Let \emph{Plain} attack refer to the case without using our proposed structures. Table \ref{result-vs-plain} gives the ACC results of Plain attack and our ambiguity attacks using CERB and IERB. For the simple AlexNet, the performance of our attack with CERB overwhelms that of Plain attack, and the performance gain becomes more significant when smaller number of training data is available. For instance, when 2\% training data are used, the ACC gap is almost 20\%. Such a property is very valuable in practice, as attacks are usually launched with very limited number of data; otherwise, re-training an entire model could be feasible as well. Also, for this simple AlexNet, our attack with IERB has similar performance with Plain attack. For the relatively large network ResNet18, our attack with CERB achieves 4.93\% ACC drop, compared with the original accuracy, when only 2\% training data are used. This drop shrinks to 1.79\% when 10\% training data are available. Compared with Plain attack, the ACC gains of our attack with CERB can be as large as 16.21\%. In addition, it is noted that our attack with IERB becomes much superior to Plain attack, when very limited training data are adopted; the ACC gain can be up to 14.53\%. More results on more complicated datasets including Caltech-101 and Caltech-256 can be found in Table \ref{result-caltech}, and similar conclusions can be drawn.                         

\begin{table*}[htb]
	\centering
	\scalebox{0.9}{
	\begin{tabular}{c|ccccccc}
		\toprule
		\cmidrule{1-8}
		WM method & Dataset-Net & Embed-Position & DataSize & Original ACC & Attack ACC   & BDR & SDR\\
		\midrule
		Greedy-Residual & Caltech256-ResNet18 & First Conv layer & 10\% & 54.98 & 54.37 & 49.22 & 100 \\
        DeepSigns & CIFAR10-WRN & Last Linear layer& 10\% & 91.30 & 91.15 & 37.5 & 100 \\
        Uchida & CIFAR10-WRN & Third Conv layer & 10\% & 90.11 & 89.77 & 45.31 & 100 \\
        
		\bottomrule
	\end{tabular}}
	\caption{Results of our ambiguity attack on other DNN watermark methods, in terms of ACC (\%), BDR (\%), and SDR (\%).}
	\vspace{-3mm}
	\label{attack-other-wm}
\end{table*}

\subsection{Attack with non-overlapping dataset}
\label{OOD-section}

In the above experiments, we consider the case that the attacker has access to a part of the original training data, namely, the dataset for launching the attack overlaps with the original one. We now investigate the attack performance under a more challenging scenario, i.e., the dataset available to the attacker and the original dataset come from the same source; but do not overlap. This non-overlapping dataset mimics the practical scenario that the attacker may not be able to exactly access a part of the original training data; but rather can only access some similar ones. 

We randomly divide CIFAR10 into non-overlapping two subsets: 40000 for training the passport-based network and the remaining 10000 for the attack. Specifically, the attacking model is a ResNet18 with 10 passport layers, and the accuracy of the trained model is 93.12\%. The attack performance of our ambiguity attacks with IERB and CERB under this new setting is tabulated in Table \ref{ood}. With these 10000 non-overlapping data, our ambiguity attack with CERB reaches a 92.17\% accuracy, only 0.95\% away from the original one. Also, the BDR in this case is as high as 91.77\%, indicating the recovered scale factors are very different from the authorized ones. When less number of non-overlapping data are used, the ACC values drop while the accompanied BDRs tend to improve. Even when only 5\% non-overlapping data are available, the ACC value can still be 85.67\% with BDR being 93.44\%. Similar results can be obtained by our ambiguity attack with IERB, but with a much lowered BDR of around 40\%. Another interesting phenomenon is that the ambiguity attack with CEBR using non-overlapping dataset leads to slightly worse ACC, but much better BDR performance, compared with the same attack with overlapping dataset. These results, again, show that our proposed ambiguity attacks are still very effective even in the challenging non-overlapping scenarios. 

\begin{table}[htb]
\centering
\scalebox{0.75}{
\begin{tabular}{c|c|c|c|c|c|c}
	\toprule
	\cmidrule{1-7}
	\# of CERB & Original & 0 & 4 & 6 & 8 & 10 \\
	\midrule
	ACC (\%) & 55.42 & 25.03 & 28.99 & 38.50 & 42.58 & 47.07 \\
	\bottomrule
\end{tabular}}
\caption{Results of using different number of CERB on Caltech-256 with ResNet18.}
\vspace{-5mm}
\label{number-of-fc}
\end{table}

\subsection{Result of ambiguity attack with CERB on other DNN watermark methods}
\label{other-methods-section}

We now show that our proposed attack strategy can be generalized well to other DNN watermark methods. As the ambiguity attack with CERB overwhelms the one with IEBR, we adopt it in the following evaluations. The three DNN watermarking methods considered are: Greedy-Residual\cite{liu2021watermarking}, DeepSigns\cite{darvish2019deepsigns}, and Uchida\cite{uchida2017embedding}, which used specific layer weights for the watermark embedding. For Greedy-Residual, the watermark was embedded into the first convolutional layer weight of the network, while for DeepSigns, the watermark was hidden in the flattened features before the last linear layer. Regarding Uchida's method, the watermark was injected into the third convolutional layer weights of the network.        

We flexibly adapt our CERB structure for different types of embedded intermediaries such as model weights and features in three DNN watermark methods. Following the attack setting of Fan \etal \cite{9454280}, we assume that the network structure and model weights except those from the watermark are available to the attacker. As a result, for Greedy-Residual, we can replace the normalization layer after the embedded convolutional parameters with the CERB. Similarly, for DeepSigns, the normalization layer before the embedded feature map is replaced by our CERB. For Uchida's method, a CERB structure is added after the embedded convolutional parameters. We preset a different signature and only train the parameters embedded with the watermarks and the CERB parameters to match this new signature. More implementation details can be found in Appendix A. 

We introduce signature detection rate (SDR) to evaluate if a signature is successfully embedded in the model. The SDR is defined as the percentage of the watermark bits $\mathbf{wm}=\left\{wm_1, ..., wm_C\right\}$ extracted from the model that are coincident with the signature $\mathbf{sig}=\left\{sig_1, ..., sig_C\right\}$ used during the embedding process, i.e., 

\vspace{-2mm}

\begin{equation}
    SDR = \dfrac{1}{C}{\sum\limits_{i=1}\limits^{C}}\mathcal{I}\Big[wm_i = sig_i \Big].
\end{equation}

The attack results on three watermark methods are presented in Table \ref{attack-other-wm}. For Greedy-Residual method, the inference performance of the model after the attack is quite similar to that of the original one, with a very slight drop of 0.61\%. In addition, the SDR is 100\%, implying that the attacker can claim the ownership to this model with the new signature. Meanwhile, a high BDR of 49.22\% well demonstrates the high dissimilarity between the forged and original watermarks. Very similar observations can be made when attacking DeepSigns and Uchida’s methods. Therefore, we validate the generalization capability of our attack strategy to other DNN watermark methods.           

\subsection{Ablation study}
\label{ablation}

\textbf{Different number of CERB structures}: To study the effect on attack performance with different number of CERB structures, we use a ResNet18 with the last 10 layers embedded with passports for the illustration. As can be seen from Table \ref{number-of-fc}, the original performance evaluated on Caltech-256 is 55.42\%. For Plain attack with 0 CERB structure, the ACC is very low, i.e., 25.05\%.  By gradually applying more CERB structures, the ACC values improve constantly. Eventually, with all 10 passport layers replaced by our CERB structures, the ACC of our ambiguity attack reaches 47.07\%.            

\textbf{Increasing the depth of CERB}: The CERB in the default setting is a two-layer perceptron with LeakyReLU. We now try to testify if using perceptron with more layers leads to better attack performance. To this end, we experiment on the ResNet18 with 10 passport layers trained on CIFAR10, and launch the ambiguity attacks by varying the number of layers in CERB. As can be seen from Table \ref{num-of-MLP}, the ACC restored by CERB with 2 layers is the highest, reaching 92.24\%. With more layers being involved, the ACC performance actually drops surprisingly. We conjecture that the two-layer perceptron is enough for our ambiguity attack; while increasing the number of layers places heavier burden to the training and eventually affects the performance of the ambiguity attack.

\begin{table}[!tbp]
	\centering
    \scalebox{0.75}{
	\begin{tabular}{c|c|c|c|c|c}
		\toprule
		\cmidrule{1-6}
		\# of MLP layers in CERB & Original & 2 & 3 & 4 & 5 \\
		\midrule
		ACC (\%) & 94.40 & 92.24 & 86.85 & 84.18 & 86.45 \\
		\bottomrule
	\end{tabular}}
	\caption{Results of changing the number of layer in CERB.}
	\label{num-of-MLP}
\end{table}

\textbf{Using other activation functions in CERB}: We also evaluate the impact of the activation functions in CERB on the overall attack performance. Again, we  experiment on the ResNet18 with 10 passport layers trained on CIFAR10. We then replace every passport layer with CERB, and adopt different activation functions in CERB. We list the ACC results by using tanh and Sigmoid, in addition to our default LeakyReLU in Table \ref{result-of-activations}. As can be noticed, all ACC results with different activation functions are similar. This implies that the attack performance is not sensitive to the activation functions adopted.            

\begin{table}[!tbp]
	\centering
	\scalebox{0.9}{
	\begin{tabular}{c|ccc}
		\toprule
		\cmidrule{1-4}
		Activation Functions & LeakyReLU & tanh & Sigmoid \\
		\midrule
		ACC (\%) & 92.87 & 92.54 & 92.80 \\
		\bottomrule
	\end{tabular}}
	\caption{Result of using different activation functions.}
	\label{result-of-activations}
        \vspace{-5mm}
\end{table}

\section{Conclusion}
	
In this paper, we propose an advanced ambiguity attack that defeats the passport-based model IP protection scheme. We combine multi-layer perceptron with skip connection to find valid substitute passports using less than 10\% of the training dataset. Extensive experimental results validate the effectiveness of our ambiguity attack. Further, it is demonstrated that our attack strategy can be easily extended to other DNN watermark methods.   

Regarding the remedy solutions, one potential direction is to exploit random locations for inserting the passport-layers, in which the randomness is controlled by a secret key. Additionally, another promising attempt is to change the embedding position from model weights to the activations. To make sure the next convolutional layer can extract proper features from the activation, the statistic of the activation should stay in a restricted scope. Such a statistic could be unique for a given signature, which may be helpful to resist our ambiguity attack.  

\textbf{Acknowledgments}: This work was supported in part by Macau Science and Technology Development Fund under SKLIOTSC-2021-2023, 0072/2020/AMJ, and 0022/2022/A1; in part by Research Committee at University of Macau under MYRG2020-00101-FST and MYRG2022-00152-FST; in part by Natural Science Foundation of China under 61971476; and in part by Alibaba Group through Alibaba Innovative Research Program.

{\small
\bibliographystyle{ieee_fullname}
\bibliography{egbib}
}

\end{document}